\documentclass[aps,twocolumn,pra,prl,superscriptaddress,preprintnumbers,amsmath,amssymb,nofootinbib]{revtex4}
\usepackage{graphicx}
\usepackage{color}
\usepackage{braket}
\usepackage{comment}
\usepackage{hyperref}

\usepackage{tikz}
\usepackage{mathtools}
\usepackage{physics}
\usepackage{simplewick}
\usepackage[normalem]{ulem}
\usepackage[obeyFinal]{todonotes}

\definecolor{refkey}{gray}{0.45}
\definecolor{labelkey}{RGB}{155,48,48}
\definecolor{UI_blue}{RGB}{32, 64, 151}
\definecolor{UI_red}{RGB}{187, 62, 24}
\definecolor{UI_blue2}{RGB}{0, 84, 147}
\definecolor{UI_red2}{RGB}{159, 32, 66}
\definecolor{UI_gray}{RGB}{169, 169, 169}
\definecolor{UI_sepia}{RGB}{112, 66, 20}
\definecolor{UI_bittersweet}{RGB}{254, 111, 94}
\definecolor{UI_emerald}{RGB}{80, 200, 120}
\definecolor{UI_olivegreen}{RGB}{181, 179, 92}
\definecolor{UI_cadetblue}{RGB}{95, 158, 160}
\definecolor{UI_fuchsia}{RGB}{255, 0, 255}
\definecolor{UI_midnightblue}{RGB}{25, 25, 112}
\definecolor{UI_royalblue}{RGB}{0,35, 102}
\definecolor{UI_periwinkle}{RGB}{204, 204, 255}
\definecolor{UI_redorange}{RGB}{255, 83, 73}
\definecolor{UI_brickred}{RGB}{203,65,84}	
\definecolor{UI_forestgreen}{RGB}{34, 139, 34}
\definecolor{UI_tan}{RGB}{210,180,140}	
\definecolor{UI_burlywood}{RGB}{222,184,135}
\definecolor{UI_burlywood}{RGB}{192,64,0}
\definecolor{UI_darkorchid}{RGB}{153,50,204}

\def\Tr{\hbox{Tr}}
\newcommand{\be}{\begin{equation}}
\newcommand{\ee}{\end{equation}}
\newcommand{\bea}{\begin{eqnarray}}
\newcommand{\eea}{\end{eqnarray}}

\newcommand{\half}{\frac{1}{2}}

\def\order{\ensuremath{\mathcal{O}}}

\begin{document}
\preprint{OU-HET-1187}

\title{The local SYK model and its triple-scaling limit}

 \author{Takanori Anegawa}\email[]{takanegawa@gmail.com}
 \author{Norihiro Iizuka}\email[]{iizuka@phys.sci.osaka-u.ac.jp} 
 \author{Sunil Kumar Sake}\email[]{sunilsake1@gmail.com}

\affiliation{Department of Physics, Osaka University, Toyonaka, Osaka 560-0043, JAPAN}

\date{\today}

\begin{abstract}
We study a model of fermions with random couplings similar to conventional SYK with $N$ number of flavours of fermions, at large $N$.
Unlike the conventional SYK model, which has all-to-all couplings, the model we study, which we call local SYK,  has a much less number of random couplings, just $N$ in number and with only local interactions. It is shown that there exists a limit in which the local SYK model can be solved using the chord diagram techniques, analogous to the double-scaled limit of conventional SYK. This limit corresponds to taking the size of the fermion coupling terms, $q$, to scale linearly with $N$. A further triple-scaling limit is taken to analyze the low energy limit and it is shown that the OTOCs saturate the chaos bound, paralleling the analysis in the conventional SYK. 
\end{abstract}

\maketitle

\setcounter{footnote}{0}

\section{Introduction}
\vspace{-1mm}

The SYK model \cite{Sachdev:1992fk, Kitaev-talks:2015}, a quantum mechanical model of fermions with random couplings, has garnered a lot of attention recently. 
The emergence of a conformal symmetry in the low energy theory and the pattern of the symmetry breaking gives rise to a (pseudo) Goldstone mode with  a Schwarzian action. A similar pattern of symmetry breaking and the associated Goldstone mode with a  Schwarzian action arises in the two dimensional JT gravity, and this leads to the  correspondence between the SYK model and JT gravity in the low energy limit\cite{Maldacena:2016hyu,Polchinski:2016xgd}. 
Another non-trivial matching between the SYK model and the bulk JT gravity 
is the growth of the out-of-time-ordered correlators (OTOC) for matter fields which saturate the maximal Lyapunov bound indicating that the SYK model is a chaotic model \cite{Maldacena:2015waa}.

A natural question that arises is about the generality of the low-energy sector of the SYK model. 
To this end, different variants of the model have been studied, such as the supersymmetric version \cite{Fu:2016vas}, models with reduced randomness \cite{Xu:2020shn}. For SYK models, typically the low energy analysis is done by a saddle point approximation in the large $N$ limit, where $N$ is the number of flavours of fermions. However, the saddle point expansion breaks down for large $q$ ($q> \sqrt{N}$), where $q$ is the size of the interactions in the Hamiltonian (see appendix A of \cite{,Anegawa:2023vxq}); $H \sim \psi^q$. The limit of $q\sim \sqrt{N}$ is the so-called double-scaling limit. In this limit, chord diagram technique \cite{Erdos:2014zgc, Berkooz:2016cvq} is suited for the analysis.  It has also been shown that the double-scaled SYK model in the low-energy limit, also leads to a Schwarzian theory\cite{Lin:2022rbf} which in turn means that the Lyapunov exponent is again maximal.

Another interesting variation of these models is regarding the randomness. The original SYK model had ${}_NC_q$ number of random couplings. 
It is important to understand if the interesting properties of the low energy spectrum of  SYK model survive as we vary the number of random couplings. A version of this question has been studied in \cite{Xu:2020shn} as sparse SYK models. In this work, the random couplings were made probabilistic, {\it i.e.}, turned on or off with some probability,  and this way, the effective number of couplings were made to vary. 
If $p$ is the probability for any of the couplings to be turned on,  (which we take to be the same for all the couplings),
then the effective number of couplings is $N_{\text{eff}}= p\times {}_NC_q $. 
It was shown that as long as the effective number of couplings $N_{\text{eff}}$  is sufficiently large, more precisely $N_{\text{eff}}=N^\alpha$ with $\alpha>1$, the model has the same saddle point in the low energy limit as the full SYK model.  For $\alpha\leq 1$, the saddle point equations change drastically and become more complicated to solve. This means that if the number of random couplings in the SYK model is reduced to $\order{(N)}$, the saddle point is expected to change \cite{Xu:2020shn,Anegawa:2023vxq}.

An analysis of a model with $\order(N)$ couplings is not explored analytically as far as we are aware. Our plan in this short paper is to consider one such model of SYK with just $N$ random couplings and analyze it. We consider the following model;
	\begin{align}
			H=i^{\frac{q}{2}}\sum_{i=1}^N J_{i , i+1 , \dots , {i+q-1}}\psi_{i}\psi_{i+1}\dots \psi_{i+q-1}\label{hrnco}
	\end{align}
where $J_{i , i+1 , \dots , {i+q-1}}$ are the random couplings. Note that the indices in $J_{i , i+1 , \dots , {i+q-1}}$ are continuous from $i$ to $i + q-1$. Imagine that we arrange the $N$ fermions on a circle. Then 
the interaction term in eq.~\eqref{hrnco} can be thought of as the coupling of fermions with only nearest-neighbor interactions  ranging from $\psi_i$ to $\psi_{i + q-1}$.  As can be seen from the above Hamiltonian there are only $N$ number of random couplings. We call this model as {\it local} SYK model  to contrast with the conventional SYK model. 

The conventional SYK has a good large $N$ limit in which  the melon diagrams dominate. However, there is no good large $N$ saddle point for the local SYK model for any fixed value of $q$, as can be easily seen from the Feynman diagrammatics. 
Hence, one approach is to rely on numerical analysis \cite{Anegawa:2023vxq}. Another possibility which we explore in this paper is to consider a large $q$ limit, along with large $N$. In fact, we find that the model is amenable to an analytical analysis in this limit. In particular, we explore a triple-scaling limit of this model and we find that an appropriate such limit is given by 
\begin{align}
	q \to \infty \,, \quad N\rightarrow \infty\,, \quad \frac{q}{N}=\text{fixed}{\,\,\ll 1}.\label{dblsyk}
\end{align}

The main result of our work is to show that there exists a good {triple} scaling limit mentioned above and the chord-diagram technique used for the fully random SYK model carries over to our model in a straighforward fashion. Furthermore, restricting to low energies,
 we find a Schwarzian-like behaviour.  In particular, we show, by considering matter operators which schematically are of the form $M\sim \psi^q$ with different set of random couplings, that the OTOC of such operators has a Lyapunov exponents which saturates the chaos bound \cite{Maldacena:2015waa}.  {The reason for our emphasis on the triple-scaling limit will be clear when we discuss the chord diagram technique in the next section.} 


\section{{Double Scaling limit and} the Chord diagrams}
\vspace{-3mm}

To begin with, let us review the method of chord-diagrams to compute the partition function in conventional SYK \cite{Berkooz:2016cvq}. This involves computing the moments of the Hamiltonian by expanding the partition function as
\begin{align}
	Z=\langle \Tr \left[ e^{-\beta H}\right] \rangle_J = \sum_k\bigg\langle \Tr\left[ \frac{(-\beta H)^k}{k!}\right ]\bigg\rangle_J \equiv    \sum_k m_k
\end{align}
where 
\begin{align}
m_k=\frac{(-\beta)^k}{k!} \langle  \Tr \left(H^k\right) \rangle_J \,,
\label{nthterm}
\end{align}
and $\langle \dots \rangle_J$ indicates that the quantities are disorder-averaged over the random couplings $J$. 
For the evaluation of this term, we think of this term as $k$ copies of Hamiltonian on a circle, corresponding to trace. Each Hamiltonian term in eq.~\eqref{nthterm} has ${}_NC_q$ random couplings. Consider one particular choice of $k$ random couplings, one from each Hamiltonian.  Since the distribution for the random coupling $J$'s are Gaussian, we perform Wick contractions among the random coupling $J$'s for the particular choice of couplings under consideration. This contraction among the $k$ couplings can be represented in the form of a diagram with chords, where a chord between two points represents the contraction between the associated random couplings. Thus, the evaluation of the term in eq.~\eqref{nthterm} entails the evaluation of various such chord diagrams corresponding to different choice of $k$ random couplings and different ways of contractions among them.

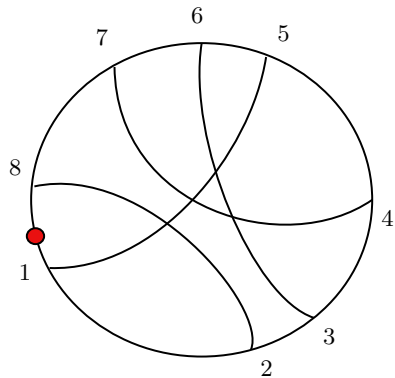
\begin{figure}[h!]
	

\tikzset{every picture/.style={line width=0.75pt}} 

\begin{tikzpicture}[x=0.75pt,y=0.75pt,yscale=-0.9,xscale=0.9]
	
	\draw   (216.88,117.45) .. controls (216.88,68.91) and (259.67,29.56) .. (312.45,29.56) .. controls (365.23,29.56) and (408.01,68.91) .. (408.01,117.45) .. controls (408.01,166) and (365.23,205.35) .. (312.45,205.35) .. controls (259.67,205.35) and (216.88,166) .. (216.88,117.45) -- cycle ;
	\draw    (263.6,42.95) .. controls (264.82,123.31) and (359.47,150.94) .. (408.01,117.45) ;
	\draw    (348.55,37.37) .. controls (340.05,92.06) and (286.66,157.91) .. (227.2,155.68) ;
	\draw    (218.7,109.92) .. controls (281.81,96.53) and (349.76,178.01) .. (340.05,201.44) ;
	\draw    (312.45,29.56) .. controls (303.95,84.25) and (341.27,172.42) .. (375.25,183.59) ;
	
	\draw (208.35,152.2) node [anchor=north west][inner sep=0.75pt]   [align=left] {1};
	\draw (343.66,205.78) node [anchor=north west][inner sep=0.75pt]   [align=left] {2};
	\draw (378.85,187.92) node [anchor=north west][inner sep=0.75pt]   [align=left] {3};
	\draw (411.61,121.79) node [anchor=north west][inner sep=0.75pt]   [align=left] {4};
	\draw (353.36,18.27) node [anchor=north west][inner sep=0.75pt]   [align=left] {5};
	\draw (304.82,8.22) node [anchor=north west][inner sep=0.75pt]   [align=left] {6};
	\draw (251.43,20.5) node [anchor=north west][inner sep=0.75pt]   [align=left] {7};
	\draw (202.89,93.05) node [anchor=north west][inner sep=0.75pt]   [align=left] {8};
		\draw  [fill={rgb, 255:red, 233; green, 16; blue, 16 }  ,fill opacity=1 ] (214.46,137.82) .. controls (214.46,135.36) and (216.63,133.36) .. (219.31,133.36) .. controls (221.99,133.36) and (224.16,135.36) .. (224.16,137.82) .. controls (224.16,140.29) and (221.99,142.29) .. (219.31,142.29) .. controls (216.63,142.29) and (214.46,140.29) .. (214.46,137.82) -- cycle ;
\label{FIG1}
\end{tikzpicture}
\caption{An example of Wick contraction contributing $m_8$. Wick contraction represents the $J$ contraction, where the Wick contracted indices of $J$ must be the same.}
\end{figure}

For the Wick contraction, $k$ must be positive and even integer and 
the Wick contracted indices of $J$ must be the same. This means that the fermion terms associated with this pair of random couplings must be the same. The important point is  that this diagrammatic method of calculation for $m_k$ works not only for the conventional SYK model but also for the local SYK model defined in \eqref{hrnco} as well. This is because the random variables $J$ in local SYK are also drawn from a  Gaussian distribution.  
However as can be seen from the diagram, chords in general intersect. These intersections can be understood as follows.  For contraction between the fermion terms, they must be bought next to each other. In this process, 
they have to cross other fermion terms and  pick up additional factors of $(-1)$ due to Grassmann nature of  the fermions $\psi_i$. This crossing of fermions is captured by the intersection of chords.  Let us label  a set of fermions as 
\begin{align}
	\Psi^q_I=\psi_{i_1}\psi_{i_2},\dots \psi_{i_q}, \quad I={i_1, i_2, i_3\dots i_q}\label{fermset}
\end{align}
If we anti-commute two sets of fermions $\Psi^q_{I_1}$ and $\Psi^q_{I_2}$ past each other, we will get a factor of $(-1)^f$ where $f$ is the number of common indices between $I_1$ and $I_2$. 

By taking into account these $(-1)^f$ factors, we can evaluate $m_k$ explicitly following \cite{Berkooz:2016cvq} using the chord-diagram technique. 
For that purpose, we need to evaluate 
the averaged value of this $(-1)^f$ factor, 
which turns out to be  different between conventional SYK model and local SYK model. 

\vspace{1mm}
{\it{\underline{Conventional SYK case}}}
\vspace{1mm}

For conventional SYK case, the probability of obtaining $f$ fermions in common in the two sets is given by 
\begin{align}
	p(f) = \frac{ {}_qC_f \times {}_{N-q}C_{q-f}}{ {}_NC_q}
	\label{pfex} 
\end{align}
Let us explain the above result.  First, pick $q$ fermion labels from the $N$ available fermions for the $\Psi_{I_1}$. The number of ways of doing this is ${}_NC_q$. Then we pick the fermions in the second set 
for the $\Psi_{I_2}$ where $f$ of them are common. 
For that, one has to pick $f$ fermions from $q$ fermions in $\Psi_{I_1}$ and then 
pick  the rest of the $q-f$ fermions for the second set from the $N-q$ fermions, {\it i.e.}, excluding the one that are in the first set $\Psi_{I_1}$. This yields  
  ${}_qC_f \times  {}_{N-q}C_{q-f}$ ways. 
Thus, up to an overall normalization constant  denoted as $A$, $p(f)$ is given by
\begin{align}
	p(f)=A\times\left( {}_NC_q \times {}_qC_f \times {}_{N-q}C_{q-f}\right)\label{pf}
\end{align}
The constant $A$ can be obtained by imposing the condition that probabilities sum up to unity. 
\begin{align}
	\sum_{f=0}^qp(f)=1\Rightarrow A=\frac{1}{ ({}_NC_q)^2 }\label{normpfval}
\end{align}
So, using eq.~\eqref{normpfval} in eq\eqref{pf}, we get \eqref{pfex}.

Let us now compute the appropriate scaling of $q$ with $N$ in the large $N$ limit. In the limit limit $N \gg q \gg f$, by using the Stirling's formula 
\begin{align}
	\ln n!=n\ln n -n \,,
	\label{stirl}
\end{align}
we obtain 
\begin{align}
	\ln p(f) &\approx 
	\ln(\frac{q^f}{f!} \frac{q!}{(q-f)!N!}\frac{((N-q)!)^2}{(N-2q+f)!}) \nonumber \\
	&\simeq \ln(\frac{1}{f!}) + \ln(\frac{q^{2f}}{N^f})-\frac{q^2}{N} + \mbox{(subleadings)}\label{lnterms}
\end{align}
Thus, we see that the $p(f)$  becomes
\begin{align}
	p(f)=\frac{1}{f!}\left(\frac{q^2}{N}\right)^f e^{-\frac{q^2}{N}}\label{pflqln}
\end{align}
Thus, we see from the above that the appropriate combination that appears is $q^2\over N$ and hence we need to keep this quantity fixed, which in the literature is called the double-scaling limit
\begin{align}
	q\rightarrow\infty \,, \quad N\rightarrow\infty \,, \quad \frac{q^2}{N}\equiv \frac{\lambda}{2} = \text{fixed}\label{dslimit}
\end{align}
Writing the answer in eq.~\eqref{pflqln} in terms of $\lambda$ defined in this way, we get
\begin{align}
	p(f)=\frac{1}{f!}\left(\frac{\lambda}{2}\right)^fe^{-\frac{\lambda}{2}}\label{pfinlm}
\end{align}
This is Poisson distribution. 

Now, the average value of the phase factor obtained by commuting two sets of fermions is given by 
\begin{align}
{\bf q}\equiv	\langle (-1)^f\rangle =\sum_{f}(-1)^f p(f)=e^{-\lambda}\label{phavgf}
\end{align}

\vspace{1mm}
{\it{\underline{Local SYK case}}}
\vspace{1mm}

%
	
For local SYK model, 	
the probability of obtaining $f$ fermions in common in the two sets is given by 
\begin{align}
	p(f)=\begin{cases}
		(N-2 q+1)/N,\qquad &f=0\\
		2 /N,\quad &f=1,\dots ,q-1\\
		1/N,\quad &f=q
	\end{cases}
	\label{pfforlocal}
\end{align}
We have implicitly assumed that $q<\frac{N}{2}$ in writing the above result. The reason for this is that we will be ultimately interested in the limit where $\frac{q}{N}\rightarrow 0$. 
The above result eq.~\eqref{pfforlocal} can be understood as follows. First, pick a set $q$ fermion labels (should be continuous) from the $N$ available fermions for the $\Psi_{I_1}$. The number of ways of doing this is $N$, which is the number of choices for the first fermion. Then we pick the fermions for the $\Psi_{I_2}$ where $f$ of them are common. 
For $f=0$, we have $(N-2q + 1)$ choices, and for $f=1$ till $f=q-1$, we have only two choices. Finally for $f = q$ we have unique choice.  
Thus, taking into account the overall normalization constant, we obtain \eqref{pfforlocal}.   

Similarly we can now compute $\langle (-1)^f\rangle $. We find 
\begin{align}
{\bf q}\equiv \langle (-1)^f\rangle = 
		p(0)-p(1)+p(2)\dots +p(q_1)=1-\frac{2q}{N}\, 
\end{align}
Therefore we take the following double-scaling limit in the local SYK model
\begin{align}
	q\rightarrow\infty, \quad N\rightarrow\infty, \quad \frac{q}{N}\equiv \frac{\lambda}{2} = \text{fixed}\label{dslimitlocal}
\end{align}
With this, ${\bf q}$ becomes 
\begin{align}
{\bf q}=  \langle (-1)^f\rangle  = 1 - \lambda
	\label{mflsq}
\end{align}

Let us compare the $p(f)$ and ${\bf q}$ between the case of conventional SYK case \eqref{pfinlm} \eqref{phavgf} and local SYK case \eqref{pfforlocal} and \eqref{mflsq}. They do not match. However if we take furthermore triple-scaling limit 
\be
\lambda \to 0,
\label{trplam}
\ee
{\bf q} matches in the leading order in $\lambda$. 
In fact in this paper we focus on this triple-scaling limit 
since it is only in this limit that we can obtain Schwarzian action and the  OTOC exhibit a  Lyapunov exponent saturating the chaos bound. 
%

 {Let us note one crucial difference between the double-scaling limit in conventional SYK and the local SYK. In conventional SYK, where $\lambda=\frac{q^2}{N}$ is fixed to any finite value with $N\rightarrow\infty$, the probability of triple intersections vanish and so we can treat any pair  of fermions independently of other fermion sets in the evaluation of a trace. However, this independence of intersections of fermion pairs is true in local SYK only in the triple-scaling limit where $\lambda=\frac{2q}{N}\ll 1$.  Let us elaborate this with a simple example. Consider the trace 
\begin{align}
	\Tr(\psi_{I_1}\psi_{I_2}\psi_{I_3}\psi_{I_1}\psi_{I_2}\psi_{I_3})\label{thretrc}
\end{align}
where $I_1,I_2,I_3$ are three arbitrary sets of fermions of the type mentioned in eq.~\eqref{hrnco}.
Naively, we would assign a factor of $\bf{q}^3$ for this term for the three exchanges of fermion sets. An explicit evaluation of this can be done  by computing 
\begin{align}
	&	\sum_{m,n,r}(-1)^{m+n+r}p(m,n,r)\nonumber\\
	&	m=|I_1\cap I_2|, \,\,	n=|I_2\cap I_3|,\,\,  	r=|I_3\cap I_1|\label{trimnr}
\end{align}
where $p(m,n,r)$ is the probability of configuration of three sets of indices $I_1,I_2,I_3$ with the number of common indices between them as in the second line of eq.~\eqref{trimnr} above. We find, in the limit of eq.~\eqref{dslimitlocal},
\begin{align}
	\sum_{m,n,r}(-1)^{m+n+r}p(m,n,r)=1-3\lambda+3\lambda^2\label{pmnr}
\end{align}
which is not exactly the same as ${\bf{q}}^3$ in the double-scaling limit. But in the triple-scaling limit in which we take $\lambda\rightarrow 0$ the above result can be approximated as $1-3\lambda\simeq {\bf q}^3$ upto corrections of $\order({\lambda^2})$ which are small and hence can be ignored.}{ In the appendix A\ref{appA}, we explicitly do the calculation of a general trace of fermion sets. We confirm that, to  leading order in $\lambda$ in the triple-scaling limit, it is consistent to use ${\bf q}$, eq.~\eqref{mflsq},  for exchange of any pair of fermions sets as if they independent. }{Hence the technique of chord diagrams only works in the triple-scaling limit for the local SYK model.}

\section{Strength of Random couplings}
\vspace{-3mm}
	
Having obtained the factor {\bf q} for the intersection of two chords in the chord diagram, let us now consider the evaluation of the term eq.~\eqref{nthterm}. To obtain a value for this term we need to determine the value of the two point function of the random variable $J$. Let us parametrize this two-point function in terms of $\sigma$ as
\begin{align}
	\langle J_{I_1}J_{I_2}\rangle =J^2\sigma\,\,\delta_{I_1 I_2}\label{jsigm}
\end{align}
The $N$-scaling of $\sigma$ in the two point function for the random variable $J$ above,  
can be fixed by the requirement that the leading term in $m_k$ is independent of $N$ as we show below. Let us elaborate on this in both conventional SYK model and local SYK model.

\vspace{1mm}
{\it{\underline{Conventional SYK case}}}
\vspace{1mm}

A general term contributing to $m_k$ in eq.~\eqref{nthterm} is of the form 
\begin{align}
	\langle \Tr(H^k)\rangle =\sum_{I_1,I_2,\dots I_k}\langle J_{I_1}J_{I_2}\dots J_{I_k}\rangle \Tr(\Psi_{I_1}\Psi_{I_2}\dots \Psi_{I_k})\label{trhinpsj}
\end{align}
where $k$ is even and we use the notation
\begin{align}
	I=\{i_1\dots i_q\},\,\,\Psi_I=\psi_{i_1}\psi_{i_2}\dots \psi_{i_q}\label{ishont}
\end{align}
Using the two point function to compute eq.~\eqref{trhinpsj}, we find
\begin{align}
	\langle \Tr(H^k)\rangle = \order\left({	(J^2\sigma)^{\frac{k}{2}}\left({}_NC_q\right)^{\frac{k}{2}} }\right)+\dots \,,
	\label{j2kdico}
\end{align}
The factor  $(J^2\sigma)^{\frac{k}{2}}$ is simply due to the contractions of $J$'s in eq.~\eqref{trhinpsj} in pairs using eq.~\eqref{jsigm}.
The factor of $\left({}_NC_q\right)^{\frac{k}{2}}$ is because there are ${}_NC_q$ choices for indices $I$ in each $J_I$ and the exponent $\frac{k}{2}$ is appropriate to the case when all the $\frac{k}{2}$ pairs are distinct pairs. The function $F(k,{\bf q})$ is due to the combinatorics of chord diagrams and the intersection of chords and this does not scale with $N$ in the double-scaling limit. The dots denote the subleading terms which are suppressed by factors of ${}_NC_q$ compared to the leading term, which happens when not all of the $\frac{k}{2}$ pairs of $J$'s are distinct. 
Thus, if we choose $\sigma$ to be such that
\begin{align}
	\sigma\sim \order{\left(\frac{1}{{}_NC_q}\right)} \,.
	\label{sinlsyk}
\end{align}
then in the large $N$ limit, we can obtain a finite answer, with only the leading term in eq.~\eqref{j2kdico} giving a non-vanishing contribution.

\vspace{1mm}
{\it{\underline{Local SYK case}}}
\vspace{1mm}

In local SYK, the $N$-scaling of the random variable $J$ is different from the conventional SYK. To understand this, we can now run the same argument as 
eq.~\eqref{sinlsyk}. In the local SYK case, 
the analog of eq.~\eqref{j2kdico}, is given by 
\begin{align}
		\langle \Tr(H^k)\rangle = \order\left((J^2\sigma)^{\frac{k}{2}}N^{\frac{k}{2}} \right)+\dots
		\label{j2mdicolsy}
\end{align}
since there are $N$ choices for $I$ in each $J_I$ to contract in local SYK model. 
Thus, the appropriate scaling of the $\sigma$ in this case is given by 
\begin{align}
	\sigma\sim \order{(N^{-1})} \,.
	\label{silsyk}
\end{align}

\section{Triple Scaling Limit and Schwarzian Action}
\vspace{-3mm}

With all the pieces in place, let us now discuss the computation of the moments eq.~\eqref{nthterm}. From these results, we can further take a limit, called the triple-scaled limit, where we obtain the Schwarzian action. 
As mentioned earlier, possible contractions in eq.~\eqref{nthterm} can be represented as possible ways of drawing chords with $k$ insertions on the circle, with one such configuration shown in FIG. 1.
We cut the circle at any point (red dot in the circle in FIG. 1, the choice of where to cut the circle will be irrelevant) turning the circle into a straight line, see FIG. 2 below. 
\begin{figure}[h!]
	
	\label{chsykex}

\tikzset{every picture/.style={line width=0.75pt}} 

\begin{tikzpicture}[x=0.75pt,y=0.75pt,yscale=-0.9,xscale=0.9]
	
		\draw (168,328) node [anchor=north west][inner sep=0.75pt]   [align=left] {1};
	\draw (203.5,329) node [anchor=north west][inner sep=0.75pt]   [align=left] {2};
	\draw (244.5,330) node [anchor=north west][inner sep=0.75pt]   [align=left] {3};
	\draw (285.5,328.75) node [anchor=north west][inner sep=0.75pt]   [align=left] {4};
	\draw (325.5,329) node [anchor=north west][inner sep=0.75pt]   [align=left] {5};
	\draw (366.5,329) node [anchor=north west][inner sep=0.75pt]   [align=left] {6};
	\draw (404.5,329) node [anchor=north west][inner sep=0.75pt]   [align=left] {7};
	\draw (439,328) node [anchor=north west][inner sep=0.75pt]   [align=left] {8};

	\draw    (154.5,323.75) -- (463.5,325) ;
	\draw  [fill={rgb, 255:red, 233; green, 16; blue, 16 }  ,fill opacity=1 ] (145.5,323.75) .. controls (145.5,321.26) and (147.51,319.25) .. (150,319.25) .. controls (152.49,319.25) and (154.5,321.26) .. (154.5,323.75) .. controls (154.5,326.24) and (152.49,328.25) .. (150,328.25) .. controls (147.51,328.25) and (145.5,326.24) .. (145.5,323.75) -- cycle ;
	\draw  [fill={rgb, 255:red, 233; green, 16; blue, 16 }  ,fill opacity=1 ] (459.5,325) .. controls (459.5,322.79) and (461.29,321) .. (463.5,321) .. controls (465.71,321) and (467.5,322.79) .. (467.5,325) .. controls (467.5,327.21) and (465.71,329) .. (463.5,329) .. controls (461.29,329) and (459.5,327.21) .. (459.5,325) -- cycle ;
	\draw    (329.5,324) -- (329.5,271) -- (171.5,271) -- (172,323) ;
	\draw    (444.5,324) -- (445.5,257) -- (209.5,257) -- (209.5,323) ;
	\draw    (371.5,325) -- (371.5,283) -- (250.5,282) -- (251,323) ;
	\draw    (410.5,325) -- (410.5,290) -- (289.5,289) -- (290,324) ;

\end{tikzpicture}
\caption{FIG. 1 is cut at red circle and is opened.}
\end{figure}
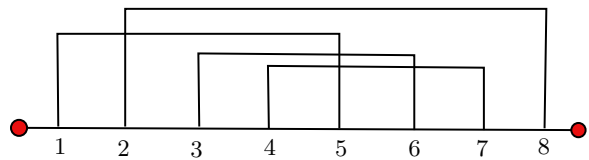

All the contractions will now be enclosed in the region between the two cuts. So, there are no chords before the first node (the node immediately after the cut = the left red circle) or after the last node (the node just before the cut = the right red circle). We now evaluate the quantity eq.~\eqref{nthterm} in the auxiliary Hilbert space called the chord Hilbert space, which is just the Hilbert space of state labelled by a non-negative integer, labelling the number of open chords. So, the quantity in eq.~\eqref{nthterm} is an transition element from $|0\rangle$, the zero chord state to $\langle 0| $ to the final zero chord state.  Let $v^{i}$ be the number of chords of open chords after $i$ nodes. As shown in \cite{Berkooz:2016cvq}, the quantity $v^{i}$ follows a recursion relation which can be written as 
\begin{align}
	v^{(i+1)}=Tv^{(i)}\label{recvrel}
\end{align}
where $T$, the transfer matrix, is given in the chord basis by 
\begin{align}
	T=\left(\begin{matrix}
		&0 & \frac{1-{\bf q}}{1-{\bf q}}& 0 &0&\cdots\\
		&1 & 0& \frac{1-{\bf q}^2}{1-{\bf q}}& 0 &\cdots\\
		&0 &1&0 & \frac{1-{\bf q}^3}{1-{\bf q}}&\cdots\\
		&0 &0 &1&0 &\cdots\\
		&\vdots &\vdots &\vdots&\vdots& \ddots
	\end{matrix}
	\right)\label{T}
\end{align}
In obtaining the above matrix for $T$, the inner product for the chord number states is taken to be
\begin{align}
	\langle m|n\rangle =\delta_{m,n}\label{delmn}
\end{align}
Thus the answer for $m_k$ can immediately be written as
\begin{align}
	m_k=\langle 0|T^{k}|0\rangle \label{mkinTk}
\end{align}
It then follows that 
\begin{align}
	\Tr(e^{-\beta H})=\langle 0 |e^{-\beta T}|0\rangle\label{trinT}
\end{align}
The operator $T$ can be written in terms of chord creation and annihilation operators $\alpha,\alpha^\dagger$ defined by 
\begin{align}
	|n+1\rangle =\alpha^\dagger |n\rangle ,\quad |n-1\rangle =\alpha|n\rangle \label{chocrean}
\end{align}
and thus $T$ becomes
\begin{align}
	T=\alpha^\dagger +\alpha\, {W} \label{Tinalalpd}
\end{align}
where $W$ is an operator acting on chord number states as
\begin{align}
	W|n\rangle =W_n|n\rangle,\quad W_n=\frac{1-{\bf q}^n}{1-{\bf q}}\label{wn}
\end{align}
It is easy to check that 
\begin{align}
	\frac{\langle n+1|T|n\rangle}{{||n+1||\, ||n||}}=1,\,\,	\frac{\langle n-1|T|n\rangle}{{||n-1||\, ||n||}}=W_n\label{tinpr}
\end{align}
However, since $T$ is not real symmetric, it won't be a Hermitian matrix. So, to remedy this and make $T$ symmetric,  we renormalize the states as follows. Let $\tilde{n}$ be the renormalized chord number state defined as
\begin{align}
	|\tilde{n}\rangle =\sqrt{S_n}|n\rangle\label{ntninde}
\end{align}
where
\begin{align}
	\quad S_n=\prod_{i=1}^{n}\frac{1-{\bf q}^i}{1-{\bf q}}, \,\, S_0=1\label{sm}
\end{align}
The annihilation and creation operators are appropriately redefined as
\begin{align}
	\tilde{\alpha}^\dagger=\sqrt{W}\alpha^\dagger, \,\,\tilde{\alpha}=\alpha \sqrt{W^{-1}}\label{alphtildef}
\end{align}
in terms of which 
\begin{align}
	\tilde{\alpha}^\dagger |\tilde{n}\rangle =|\widetilde{n+1}\rangle,\,	\tilde{\alpha} |\tilde{n}\rangle =|\widetilde{n-1}\rangle
\end{align}
with the renormalized $T$ given by 
\begin{align}
	\tilde{T}=\tilde{\alpha}^\dagger + \tilde{\alpha}\,W \label{ttilde}
\end{align}
Written in terms of the original operators $\alpha, \alpha^\dagger$, it reads
\begin{align}
	\tilde{T}=\sqrt{W}\,\alpha^\dagger +\alpha \, \sqrt{W}\label{Tsymexp}
\end{align}
Again, it is easy to check that
\begin{align}
		\frac{\langle \widetilde{n+1}|\tilde{T}|\tilde{n}\rangle}{{||\widetilde{n+1}||\, ||\tilde{n}||}}=\sqrt{W_{n+1}},\,\,	\frac{\langle \widetilde{n-1}|\tilde{T}|\tilde{n}\rangle}{{||\widetilde{n-1}||\, ||\tilde{n}||}}=\sqrt{W_n}\label{tinpr}
\end{align}
and thus the matrix of $\tilde{T}$ is symmetric. 
Rewriting in terms of the variable 
\begin{align}
	l=-n\log {\bf q}\label{link}
\end{align}
we get
\begin{align}
	\tilde{T}=\frac{1}{\sqrt{1-\bf {q}}}\left(\sqrt{1-e^{-l}}e^{-i\lambda k}+e^{i\lambda k}\sqrt{1-e^{-l}}\right)\label{Tinl}
\end{align}
where $k$ is the conjugate momentum to $l$. 
{Now considering the limit}
\begin{align}
	\lambda\rightarrow 0, \quad l\rightarrow\infty, \quad \frac{e^{-l}}{\lambda^2}=\text{fixed}\label{tripsclimi}
\end{align}
we find
\begin{align}
	\tilde{T}-E_0=-\lambda^{\frac{3}{2}}\left(k^2+{e^{-{l}}\over \lambda^2}\right),\quad E_0=\frac{2}{\sqrt{\lambda}}\label{liovact}.
\end{align}
Let us remind the reader that from eq.~\eqref{recvrel}, the operator $T$ (or $\tilde{T}$) is an evolution operator along the boundary {\it i.e.,} along the nodes after cutting the circle. Taking  large $l$ in the triple-scaling limit eq.~\eqref{tripsclimi} makes the boundary time almost continuous. 
Thus,  it can be associated with a conjugate Hamiltonian by 
\begin{align}
	\tilde{T}\propto e^{-\epsilon H_{\text{bdy}}}\label{ttashbdy}
\end{align}
where $H_{\text{bdy}}$ is conjugate to boundary time. 
Therefore we can read off $H_{\text{bdy}}$ as
\begin{align}
	H_{\text{bdy}} \propto k^2+{e^{-{l}}\over \lambda^2}\label{hbdy}
\end{align}
which is the action for the Schwarzian theory written in \cite{Lin:2022rbf}.  

Let us make a few more comments before we turn to OTOCs. Even though the chord intersection factor $\bf{q}$ is not exactly the same for the local and non-local SYK and differ at $\order{(\lambda^2)}$ in the limit of small $\lambda$, the steps leading to eq.~\eqref{liovact} are still valid in the local SYK case as well. Heuristically, the limit of $l$ satisfying eq.~\eqref{tripsclimi} can be thought of as restricting the range of  energies in the double scaled SYK to a window analogous to the case of fixed $q$ SYK where the Schwarzian theory emerges, $\frac{1}{N}\ll \frac{E}{J}\ll 1$. 

\section{OTOC}
\vspace{-3mm}
\label{lsyk}
We are interested in computing a quantity of the form 
\begin{align}
	\langle M_2(t)M_1(0)M_2(t)M_1(0)\rangle
	\label{otocm1m2}
\end{align}
where the operators $M_1,M_2$ are matter operators whose details will be discussed shortly. 
The operator $M_i(t)$ is obtained from $M_i(0)$ by doing a time evolution with the Hamitonian $H$. The matter operators, considered in conventional SYK, are also made of $\psi^q$ terms with random couplings
\begin{align}
	M=\sum_{\tilde{I}}\tilde{J}_{\tilde{I}}\psi_{\tilde{I}},\quad \tilde{I}=\{i_1,\,i_2,\,\dots i_q\}\label{matop}
\end{align}
	where $\tilde{J}$ are also random variables drawn from a Gaussian distribution.  The OTOC calculation in non-local SYK is more amenable for matter operators of the form $\psi^{q}$ where $q$ is also taken to scale as $\sqrt{N}$. Similarly, in the local SYK model under consideration, as we will see, it will be easy to analyze operators of the form 
	\begin{align}
		M_i=\sum_{\tilde{I}}\tilde{J}_{\tilde{I}}\psi_{\tilde{I}},\quad \tilde{I}=\{i_1,\,i_1+1\,\dots i_1+q_i-1\}\label{matopls}
	\end{align}
As before, we restrict ourselves  to operators of sizes $q_i$ such that
\begin{align}
	\frac{q_i}{N}<\half\label{qinsi}
\end{align}
The evaluation of the OTOC, eq.~\eqref{otocm1m2} requires the evaluation of terms of the form
\begin{align}
	\tilde{m}_{1234}=	\langle \Tr (M_1H^{k_4}M_2H^{k_3}M_1H^{k_2}M_2H^{k_1})\rangle \label{matcor}
\end{align}
To compute the OTOC, and in particular to evaluate terms as above eq.~\eqref{matcor}, we need to evaluate  the appropriate factors of $\langle (-1)^f\rangle$ when two fermion sets of unequal lengths cross each other. Consider two fermion sets of length $q_1, q_2$. For the conventional SYK this factor is similar to eq.~\eqref{phavgf} given by 
\begin{align}
	{\bf q_{12}}=e^{-{2q_1q_2\over N}}\label{qbfnl}
\end{align}
The computation of the analogous factor for local SYK requires a more careful analysis which we shall do now. 
 Without loss of generality we take the case $q_1\leq q_2$.  First, we pick the $q_1$ fermions out of $N$ fermions. The number of ways this can be done is $N$. Then we pick the second set of $q_2$ fermions from the set of $N$ flavours. For this set to have $f=0$ fermion in common, the number of ways of picking the second set is $(N-q_1-q_2+1)$. For $f=1$ fermion in common, the number of ways is only two, either the first or the last fermion of the first set  can be in common. So is the case till $f=q_1-1$. For the case of $f=q_1$, since $q_2>q_1$, the number of ways of picking the second set is $q_2-q_1+1$. We can now translate these into probabilities by including a proportionality factor which can be fixed by requiring probabilities add up to unity. Thus, we have
\begin{align}
	p(f)=\begin{cases}
		A N (N-q_1-q_2+1),\qquad &f=0\\
		2 A N,\quad &f=1,\dots ,q_1-1\\
		A N(q_2-q_1+1),\quad &f=q_1
	\end{cases}
	\label{pfunlsyk}
\end{align}

Requiring that probabilities add up to unity, we get
\begin{align}
	A=\frac{1}{N^2}\label{Alysk}
\end{align}


We can now compute $\langle (-1)^f\rangle $. We find 
{\footnotesize
\begin{align}
	\langle (-1)^f\rangle =\begin{cases}
		p(0)-p(1)+p(2)\dots +p(q_1)=1-\frac{2q_1}{N}\,\quad \text{for $q_1$ even}\\
		p(0)-p(1)+p(2)\dots -p(q_1)=1-\frac{2q_2}{N}\,\quad \text{for $q_1$ odd}
	\end{cases}
	\label{mflsq1q2}
\end{align}}
From the above we see the factor upon exchanges of a set of $q_1$ fermions with another set of $q_2$ fermions gives a value that depends on whether $q_1$ is even or odd. For the chord diagram technique to work, we need to restrict to the matter operators such that
\begin{align}
	{q_1\over N}=\frac{q_2}{N},\quad \text{as}\quad N\rightarrow\infty  \label{dbscq1q2}
\end{align}
We thus see that the double scaling analysis for the local SYK model imposes a significant constraint on the size of the operators. But note that this is not too constraining in the sense that the size of the operators $q_1$ and $q_2$ need not be exactly same but only that ${\abs{q_2-q_1}\over N}\rightarrow 0$ in the limit of $N\rightarrow \infty$. Thus, we have only a universal factor given by 
\begin{align}
	{\bf	\tilde{q}_l}\equiv \langle(-1)^f \rangle=1-\frac{2q}{N}\label{qtlsyk}
\end{align}
where ${\abs{q-q_1}\over N}\rightarrow 0,{\abs{q-q_2}\over N}\rightarrow 0$.
Thus, again we see that the double-scaling limit for the matter operators in the case of local SYK corresponds to 
\begin{align}
	q\to \infty \,, \quad N \to \infty \,, \quad \frac{q}{N}=\text{fixed}\label{lyskdb}.
\end{align}
Since, the OTOC calculation involves the evaluation of terms as in eq.~\eqref{matcor}, which involves the intersection of matter operators and also the Hamiltonian chords, we require
\begin{align}
	{q\over N}\equiv{q_H\over N}={q_1\over N}={q_2\over N}\label{q1q2qh}
\end{align}
where $q_H, q_1, q_2$ are the sizes of the Hamiltonian, matter operators $M_1, M_2$ respectively. 
With this understanding the OTOC can be evaluated following \cite{Berkooz:2018jqr}. The only important difference is that the factor of $(-1)^f$ is the same for the intersection of any type of chord unlike the conventional SYK. The triple-scaling limit can then be taken by taking the parameter $\lambda=\frac{2q}{N}$ to be small. We indeed get the same Lyapunov exponent as in the case of the double-scaling limit of the conventional SYK model given by 
\begin{align}
	\lambda_L=2\pi T-{4\pi\over \sqrt{\lambda}} T^2,\quad \lambda=\frac{2q}{N} \label{llsyk}
\end{align}
The above result is valid in the range of temperatures
\begin{align}
	\lambda^{\frac{3}{4}}\ll T\ll \sqrt{\lambda}\label{Tlamrel}
\end{align}
The constraint on the temperature in terms of $\lambda$ is the statement that we restrict to low energies (upper inequality) and the consistency a of saddle point analysis in deriving the result eq.~\eqref{llsyk}, see \cite{Berkooz:2018jqr}. 

Before we end this section, let us make a few comments. Although, we focused in this section on the OTOCs, one can analyze the two-point function itself which is technically simpler compared to OTOC. The analysis, { in the triple-scaling limit}, again parallels the case of conventional SYK \cite{Berkooz:2016cvq} with just the parameter ${\bf q}$ of double-scaling limit, eq.~\eqref{phavgf} of conventional SYK replaced by the parameter of local SYK, eq.~\eqref{mflsq}.

\section{Conclusion}
\label{conc}

In this paper, we considered a variant of the SYK model with reduced number of random couplings. In  particular, we considered a model with nearest neighbour fermion couplings, with the number of random couplings being $N$. We outlined how a double-scaling limit of such a model can be obtained by taking $q, N\rightarrow\infty $ with the ratio $\frac{q}{N}$ held fixed. {The triple-scaling limit of this model corresponds to further taking $\frac{q}{N}\rightarrow 0$}. We argued that the low energy limit of this  {triple} scaled local SYK model also leads to a  Schwarzian theory. This is surprising at first since the model is very sparse and would not be expected to give rise to a Schwarzian theory since it would then mean that the OTOCs would have a maximal Lyapunov exponent. A naive expectation would be that since the model is a highly sparsed version of the full SYK model, one may possibly obtain a Lyapunov growth for OTOCs which would be much smaller than the maximum value of $2\pi T$.   However, in the {double} scaled limit, the size of the fermion terms $q$ is also taken to grow proportional to $N$ thus making the model `highly' connected in some sense, which lead to the maximal Lyapunov exponent. We have also argued that, by considering appropriate matter operators, that the OTOCs indeed have the maximal Lyapunov exponent.

For the local model under consideration, the existence of a maximal Lyapunov exponent suggests that the low energy limit involves ``time-reparametrization" mode from an appropriate emergence and breaking of the conformal symmetry in the low energy limit  as happens in the conventional SYK model. It would be interesting to understand the properties of this conformal fixed point, in particular, the scaling dimension of fermion operator at this fixed point. The sparse version of the SYK model have been studied in the literature \cite{Xu:2020shn}. In \cite{Xu:2020shn}, the authors show that for fixed $q$, in the large $N$ limit, the low energy fixed point for sparse model of SYK is the same as the fully random SYK for that value of $q$. This results holds true so long as the model is sufficiently random that the effective number of random couplings is $N^\alpha$ with $\alpha>1$. This analysis can be extended to a system with $\order(N)$ effective number of random couplings but by considering the large $q$ limit \cite{Anegawa:2023vxq}. In particular, this is argued to be true even for $q\sim N^\rho$, where $0\leq \rho <\frac{1}{2}$. At $\rho=\frac{1}{2}$, the saddle point analysis in terms of $G,\Sigma$ can also be done with $\lambda=\frac{q^2}{N}\rightarrow 0$. To compare with the current investigation, the model we study has $\order{(N)}$ number of couplings with $q\sim \order{(N)}$. It would be interesting to understand the relation between the $G,\Sigma$ method for $q\sim {N}$ and contrast it the chord-diagram analysis presented above. Recall that, in the conventional SYK, the $G,\Sigma$ analysis can be done for both finite $q$ and large $q$ (even in the double-scaled regime). But for the Local SYK model, we do not expect a $G,\Sigma$ analysis to lead to a Schwarzian theory for finite $q$. However, since we find an analytically solvable limit of double scaling regime for local SYK, we can expect that a $G,\Sigma$ field analysis can also be done in this regime, where $q$ scales as $N$. 

Finally, let us make one more comment. The analysis of the local SYK model considered in this paper in the double, triple-scaling limit is specific to this model. 
For example, if we further truncate the couplings of local SYK to half by setting $J_{i,i+1\dots}$ to vanish, say when $i\mod 2=0$, the fermion couplings that are retained commute with each other. So, even in the double-scaling limit ${\bf q}=1$. Thus, the analysis leading to the Schwarzian and the derivation of the Lyapunov exponent, in the triple-scaling limit, is no longer valid. So, even though, the number of random couplings is $\order{(N)}$, more precisely in this case $\frac{N}{2}$, the conclusions are significantly different. This shows that the details depend on the specific random couplings and not just on the number of them.

\vspace{5mm}
\centerline{\bf{Acknowledgments}} 
\vspace{1mm}
\begin{acknowledgments}
We thank  Micha Berkooz, Prithvi Narayan, Onkar Parrikar, {Harshit Rajgadia}, Sandip Trivedi and  Masataka Watanabe for helpful discussions. 
We also thank Arkaprava Mukherjee and Sandip Trivedi for collaboration on  \cite{Anegawa:2023vxq}. 
The work of TA and NI were supported in part by JSPS KAKENHI Grant Number 21J20906(TA), 18K03619(NI). The work of NI and SS were also supported by MEXT KAKENHI Grant-in-Aid for Transformative Research Areas A “Extreme Universe” No. 21H05184. 
\end{acknowledgments}

\appendix
{
\section{A. Factor for exchange of fermion sets}
\label{appA}

In this section, we shall explicitly evaluate the expectation value of $(-1)$ for the exchange of arbitrary sets of fermions in the local SYK model. In this model, the factor for exchange of a pair of fermion sets is independent of other sets of fermions in the trace, in the triple-scaling limit, given by ${\bf q}$ (eq.~\eqref{mflsq}), which we shall now show\footnote{We thank Harshit Rajgadia for highlighting the importance of this issue in this model to us.}.
This justifies our emphasis on the triple-scaling limit in the local SYK model, as also mentioned around eq.~\eqref{trplam}. 

We shall now argue that the value of the trace of an arbitrary number of fermion sets in the computation of moments $m_k$ can be assigned the value, ${\bf q}^a$, where $a$ is the number of pairwise exchange of fermions needed to be done in the evaluation of the trace to annihilate them all. The error in this consideration is of $\order{(\lambda^2)}$, which is small in the triple-scaling limit.  More precisely, we argue that the value of a trace of fermion sets in the triple-scaling limit eq.~\eqref{dslimitlocal},eq.~\eqref{trplam}
\begin{align}
&	\Tr(\psi_{I_{\rho(1)}}\psi_{I_{\rho(2)}}\dots \psi_{I_{\rho(k)}}\psi_{I_{\sigma{(1)}}}\psi_{I_{\sigma{(2)}}}\dots\psi_{I_{\sigma{(k)}}})\nonumber\\
	&\quad= 1- a \lambda+\order{(\lambda^2)}\simeq {\bf q}^a\label{arbtrev}
\end{align}
where $\rho, \sigma$ are  elements of $S_k$, the permutation group of $k$ elements and $a$ is the minimum number of pairwise exchanges of fermion sets needed to be done  in eq.~\eqref{arbtrev} to annihilate them. A trace of the form in eq.~\eqref{arbtrev} written explicitly reads
\begin{align}
		\Tr(\psi_{I_1}\psi_{I_2}\dots \psi_{I_k}\psi_{I_{\sigma{(1)}}}\psi_{I_{\sigma{(2)}}}\dots\psi_{I_{\sigma{(k)}}})=\sum_{n_{ij}} (-1)^{{n}_{ij}} {\bf p}(n_{ij})\label{arbtrex}
\end{align}
The above expression requires some explanation. For exchange of two fermion sets $I_i, I_j$, we get a factor of $(-1)^{n_{ij}}$ where $n_{ij}$ is the number of common fermions between the sets $I_i, I_j$. We include such a factor for every pair which needs to be exchanged. The probability ${\bf p}(n_{ij})$ is the joint probability distribution for all the intersection numbers $n_{ij}$ that appear in the exponent of (-1).

 The point is that to compute the $\order{(\lambda)}$ term in eq.~\eqref{arbtrex}, it is enough to just consider those configuration in which the pairs whose intersection numbers appear in the exponent of $(-1)$ in eq.~\eqref{arbtrex} do not overlap, {\it i.e.}, do not have any common fermions. The contribution of the configurations in which these index sets overlap will only contribute at $\order{(\lambda^2)}$ or higher. Let us first illustrate these points with an example. We then give a proof of the above claims for more general cases. Consider the evaluation of the trace
 \begin{align}
 	\Tr(\psi_{I_3}\psi_{I_1}\psi_{I_2}\psi_{I_4}\psi_{I_1}\psi_{I_3}\psi_{I_4}\psi_{I_2})\label{tr4}
 \end{align}
The explicit expression for this is given by 
\begin{align}
	&\Tr(\psi_{I_3}\psi_{I_1}\psi_{I_2}\psi_{I_4}\psi_{I_1}\psi_{I_3}\psi_{I_4}\psi_{I_2})=\nonumber\\&\quad\sum_{n_{12},n_{14},n_{23},n_{34}}(-1)^{n_{12}+n_{14}+n_{23}+n_{34}}p(n_{12},n_{14},n_{23},n_{34})\label{tr4ex}
\end{align}
As per our claim above, the $\order{(\lambda)}$ term can be obtained by just evaluating the contribution of the configurations for which $n_{12}=n_{14}=n_{23}=n_{34}=0$. The intersection number $n_{13},n_{24}$ are unconstrained meaning that the pairs $(I_1,I_3)$ and $(I_2,I_4)$ can either overlap or be not overlap. To get the probability $p(n_{12}=0, n_{14}=0,n_{23}=0, n_{34}=0)$, we have to just count the number of configurations in which the pairs of sets $(I_1,I_2),(I_1,I_4),(I_2,I_3),(I_3,I_4)$ do not overlap. This is just number of ways of picking four sets of $q$ consecutive integers from the set of $\{0\dots N\}$(with periodicity in the sets assumed) such that the pairs mentioned have zero overlaps. Let us define the quantity
\begin{align}
	A_k(l)&\equiv\sum_{y_1=0}^{l}	\sum_{y_2=0}^{l}\dots 	\sum_{y_k=0}^{l}\delta_{y_1+y_2+y_3\dots+y_k,l}\nonumber\\
	&={}_{l+k-1}C_{k-1}\label{Akdef}
\end{align}
The last equality can be verified using induction and by noting that the quantity $A_k(l)$ satisfy the recursion relation
\begin{align}
	&A_k(l)=A_{k-1}(l)+A_{k-1}(l-1)+\dots+A_{k-1}(0)\
\end{align}
To count the number of configurations appropriate to $p(n_{12}=0,n_{14}=0,n_{23}=0,n_{34}=0)$, we consider various cyclic ordering of the index sets on the circle. There are six inequivalent cyclic orderings of the four sets on the circle. For each of the cyclic ordering, we count the number of possible configurations. The ordering and the corresponding number of configurations, in the $N,q\rightarrow \infty$ limit,  are as follows, 
\begin{align}
	&1\,2\,3\,4\Rightarrow A_4(N-4q)\nonumber\\
	&1\,2\,4\,3\Rightarrow A_4(N-2q)\nonumber\\
	&1\,3\,2\,4\Rightarrow A_4(N-2q)\nonumber\\
	&1\,3\,4\,2\Rightarrow A_4(N-2q)\nonumber\\
	&1\,4\,2\,3\Rightarrow A_4(N-2q)\nonumber\\
	&1\,4\,3\,2\Rightarrow A_4(N-4q)\nonumber\label{4setcount}
\end{align}
Combining these, we get
\begin{align}
	&\sum_{n_{12},n_{14},n_{23},n_{34}}(-1)^{n_{12}+n_{14}+n_{23}+n_{34}}p(n_{12},n_{14},n_{23},n_{34})\nonumber\\
	&\quad \simeq\frac{1}{N^3}{(2A_4(N-4q)+4A_4(N-2q))}\simeq 1-4\lambda
\end{align}
A simple rule for obtaining the numbers in eq.~\eqref{4setcount} can be obtained using the graph technique outlined below.  
\begin{figure}[h!]

	\tikzset{every picture/.style={line width=0.75pt}} 
	
	\begin{tikzpicture}[x=0.75pt,y=0.75pt,yscale=-1,xscale=1]
		
		\draw [color={rgb, 255:red, 126; green, 211; blue, 33 }  ,draw opacity=1 ][line width=3]    (233.37,203.5) -- (361.89,203.5) ;
		\draw [color={rgb, 255:red, 126; green, 211; blue, 33 }  ,draw opacity=1 ][line width=3]    (239.21,73.15) -- (367.74,73.15) ;
		\draw [color={rgb, 255:red, 126; green, 211; blue, 33 }  ,draw opacity=1 ][line width=3]    (223.63,183.7) -- (223.63,94.6) ;
		\draw [color={rgb, 255:red, 126; green, 211; blue, 33 }  ,draw opacity=1 ][line width=3]    (381.37,183.7) -- (381.37,94.6) ;
		\draw [color={rgb, 255:red, 208; green, 2; blue, 27 }  ,draw opacity=1 ][line width=3]    (363.84,190.3) -- (243.11,91.3) ;
		\draw [color={rgb, 255:red, 208; green, 2; blue, 27 }  ,draw opacity=1 ][line width=3]    (239.21,191.95) -- (361.89,88) ;
		
		\draw (217.16,194.18) node [anchor=north west][inner sep=0.75pt]   [align=left] {1};
		\draw (374.89,194.18) node [anchor=north west][inner sep=0.75pt]   [align=left] {2};
		\draw (376.84,65.47) node [anchor=north west][inner sep=0.75pt]   [align=left] {3};
		\draw (221.05,63.83) node [anchor=north west][inner sep=0.75pt]   [align=left] {4};

	\end{tikzpicture}
\caption{A graph representation of the trace in eq.~\eqref{tr4ex}}
\label{4tr}
\end{figure}
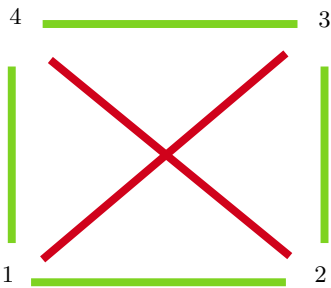

For example, the graph for the trace in eq.~\eqref{tr4} is shown in fig.\ref{4tr} As can be seen from this graph, there is an edge between every pair of vertices. The edges are of two types. The red coloured edge between $(i,j)$ vertices indicate that there is no cost of $(-1)$ when exchanging the sets $I_i, I_j$. A green coloured edge between the vertices indicate that there is a factor of $(-1)^{n_{ij}}$ between these edges. The different orderings in eq.~\eqref{4setcount} can be thought of as different closed paths on the graph, with each path passing through all the vertices once and only once. The number of configurations corresponding to a particular closed path, in the limit $N,q\rightarrow \infty$,  is given by $A_{4}(N-\alpha_i q)$, where $\alpha_i$ is the number of green edges along this path. This can be easily checked against the result in eq.~\eqref{4setcount}. We can then immediately compute the coefficient of $\lambda$ term by considering all possible inequivalent cyclic orderings. 

We can now generalize this procedure. For every trace such as eq.~\eqref{arbtrev} a graph can be constructed with coloured edges to distinguish the vertices which have a factor of $(-1)^{n_{ij}}$ for the exchange. For a trace with $k$ different fermion index sets in the trace, the leading $\lambda$ term can be obtained by the formula
\begin{align}
	\frac{1}{N^{k-1}}\sum_{i\in\text{closed paths}}A_k(N-\alpha_i q)\label{aksum}
\end{align}
Using the large argument expansion of $A_k(N-\alpha_i q)$ given by 
\begin{align}
	A_k(N-\alpha_i q)\simeq\frac{N^{k-1}}{(k-1)!}\left(1-\alpha_i (k-1)\frac{\lambda}{2}\right)\label{akexp}
\end{align}
the result in eq.~\eqref{aksum} becomes
\begin{align}
	&\frac{1}{(k-1)!}\sum_{i\in \text{closed paths}} \left(1-\alpha_i (k-1)\frac{\lambda}{2}\right)\nonumber\\
	&=1-a\lambda\label{aksumval}
\end{align}
where we used the fact that the number of closed paths is $(k-1)!$ and 
\begin{align}
\sum_{i\in \text{closed paths}}\alpha_i=	{2(k-2)!}a
\end{align}
where $a$ is the number of green edges in the graph (which is the same as the number of pairwise exchanges needed to be done to evaluate the trace). Thus, we see from eq.~\eqref{aksumval} that we indeed get the coefficient of the $\lambda$ term correctly in the triple-scaling limit by considering the configurations in which pairs with the $(-1)$ cost in the sum being disjoint. 

We are now left to argue that the remaining configurations have a contribution that is suppressed in $\order{(N^{-1})}$ or only contribute at $\order{(\lambda^2)} $ or higher in the triple-scaling limit.  Consider the contribution from configurations in which there are non-zero common fermions between two sets, (say $I_1,I_2$) which have to be exchanged in computing the trace (meaning they have a $(-1)^{n_{12}}$ cost). The number of configurations with such non-zero common fermions ($n_{12}\neq 0$) is significantly reduced as the two index sets are to overlap. The number of such configurations is in fact suppressed by a factor of $N$. If more number of such sets overlap, the number of configurations is suppressed appropriately by more powers of $N$. In the case of two overlapping sets,  since there are $\order{(q)}$ configurations, it may seem that even though there is a suppression in $N$, it is enhanced by $q$ and so we can get a factor of $\lambda$ for such configurations. However, because of the $(-1)^{n_{12}}$, there will be cancellations between the contributions of different configurations and so will not give a factor of $\lambda$. This argument can be generalized to non-zero common fermions between multiple sets. For multiple overlapping sets, the $(-1)^\#$ can be effectively nullified, but the number of such configuration will scale as a power of $q$ and so can only give a factor of $\lambda^\alpha$, where $\alpha>1$. 


Let us elaborate this point with an example.  Suppose there is a trace which has a factor $(-1)^{n_{12}+n_{23}+n_{13}}$. Consider some set of fermions which are common to the three sets $I_1,I_2,I_3$. In this case $n_{12}\neq 0, n_{13}\neq 0, n_{23}\neq 0=0$. But since the index sets are continuous integer sets, it has to be that one of the three possibilities should happen
\begin{align}
	n_{12}+n_{23}&=q+n_{13}\quad \text{or}\nonumber\\
		n_{13}+n_{23}&=q+n_{12}\quad \text{or}\nonumber\\
			n_{12}+n_{13}&=q+n_{23}
\end{align}
The contribution of the above configuration to the trace is 
\begin{align}
	\quad\sum_{n_{12},n_{13},n_{23}\dots}(-1)^{n_{12}+n_{13}+n_{23}+\dots}p(n_{12},n_{13},n_{23},\dots)\nonumber\\
	=\sum_{n_{12}, n_{14}, n_{24}\neq0}^{q}p(n_{12}, n_{13}, n_{23},\dots)\label{tirpcon}
\end{align}
This term can be evaluated explicitly and checked that it gives a contribution at $\order{(\lambda^2)}$. This completes our proof. 
}

\bibliography{refs}

\end{document}